\documentstyle[aps,prl,twocolumn,amssymb]{revtex}

\begin{document}
\title{Absolute spin-valve effect with superconducting proximity structures}
\author{Daniel Huertas-Hernando$^{1}$, Yu. V. Nazarov$^{1}$, W. Belzig$^{2}$}
\address{$^{1}$Department of Applied Physics and Delft Institute of Microelectronics
and Submicrontechnology,\\
Delft University of Technology, Lorentzweg 1, 2628 CJ Delft, The Netherlands%
\\
$^{2}$ Department of Physics and Astronomy, University of Basel,\\
Klingelbergstr. 82, 4056 Basel, Switzerland}
\date{\today}
\maketitle

\begin{abstract}
We investigate spin dependent transport in hybrid
superconductor(S)--normal-metal(N)--ferromagnet(F) structures under
conditions of proximity effect. We demonstrate the feasibility of the
absolute spin-valve effect for a certain interval of voltages in a system
consisting of two coupled tri-layer structures. Our results are also valid
for non-collinear magnetic configurations of the ferromagnets.
\end{abstract}



Spin transport in hybrid systems of ferromagnets and normal metals is a very
active field of research. This is inspired by prospectives of spin-based
electronics or ``spintronics'' \cite{RefPrinz}. The feasibility to create
and control spin accumulation in such systems by injecting spin polarized
current from a ferromagnetic material into a non-magnetic one is being
extensively studied \cite{RefJedema}. The theory predicts a variety of novel
effects in the case of non-collinear magnetizations \cite{RefArne}.

The main attention receives the so-called spin-valve effect, which provides
the mechanism for the giant magnetoresistance (GMR) \cite{Gerrit}. An
idealized ferromagnetic metal would have electrons with only one direction
of spin. The current between two such metals would not go if their
magnetizations are opposite. This is the {\em absolute spin-valve effect}.
The absolute effect is impossible to achieve with common ferromagnetic
metals, since electron states of both spin directions are present at Fermi
surface. This is why the actual values of GMR are relatively small. There
have been substantial efforts to increase these values by exploring various
material combinations \cite{Gerrit}. Recent attempts to realize the {\em \
absolute spin-valve effect} concentrated on exotic magnetic materials. A
spin polarization of up to 80 \% was achieved using the dilute magnetic
semiconductor Zn$_{1-x}$-Mn$_{x}$Se \cite{RefMolenkamp}.

In this Letter we propose a different approach, in which an {\em absolute
spin-valve effect} can be achieved without using ``exotic'' compounds. We
suggest to use the proximity effect minigap induced in a normal metal by an
adjacent superconductor. This minigap has been predicted long ago \cite
{RefMcMillan} and has been intensively investigated in recent years \cite
{RefMcMillan1}. Features related to the proximity effect can be probed by
tunneling spectroscopy measurements. The tunneling current between two
superconducting proximity structures exhibits a jump at the voltage $eV_{%
\text{th}}=(\tilde{\Delta}_{1}+\tilde{\Delta}_{2})$, $\tilde{\Delta}_{1(2)}$
being the minigaps in the structures. This is a consequence of sharp peak in
the density of states at the minigap edge, that mimics a BCS density of
states. The current jump at the threshold voltage is well known for
tunneling between superconductors \cite{RefTinkham}.

We will use the minigap to achieve an {\em absolute spin-valve effect} for
the tunneling current between two hybrid structures. Each structure combines
a normal metal part with superconducting and magnetic reservoirs, that
induce superconducting and magnetic correlations in the normal metal part.
The presence of a normal part is essential to provide a physical separation
between the sources of superconducting and ferromagnetic correlations. This
assures that neither the ferromagnet suppress superconductivity nor the
superconductor affects ferromagnetism. It also provides more control over
the strength of the correlations.

We have found that the best result is achieved if the ferromagnet is an
insulator. Then the only result of the magnetic correlations is a shift $\pm 
$ $\tilde{h}$ of the minigap edges for opposite spin directions. The peaks
of the density of states are therefore split. If one combines two such
structures by a tunnel contact between the normal metal parts, the tunneling
current exhibits jumps at different threshold voltages depending on which
spin components contribute to the current. In the voltage interval between
these threshold voltages, the tunneling current jumps from zero to a finite
value differently for parallel than for antiparallel orientations of
magnetizations in the two structures. Generally, the results depend on the
relative orientation of the magnetizations of the two ferromagnets in the
system, as well as on the induced superconducting gaps and the induced
spin-splitting in each normal metal.

A possible design for an actual device is shown in Fig.1. It consists of two
S/N/F structures as described above with their normal parts connected by a
tunnel junction. For the calculation, we adopt the circuit theory
description of the system \cite{RefYuli}. In terms of Green's functions,
this means that we assume isotropic Green's functions in momentum space.
Quasiclassical Green's functions methods have already been used to study
structures involving superconducting materials and magnetically active
interfaces \cite{RefMilis}. The advantage of the circuit theory description
is that we do not have to specify a concrete geometry of the structures.
Each part of the structure is then presented by a normal node, which is
connected to superconducting and ferromagnetic reservoirs by means of tunnel
junctions. We concentrate first on one of the structures.

In the circuit theory, the Green's functions are calculated from balance
equations for matrix ``currents'' in each node. These currents come from
each connector to the node. The matrix current is expressed in terms of the
connector properties and the Green's functions on the two sides of the
connector. The case of a matrix current, that accounts both for the
ferromagnetic and for the superconducting nature of the reservoirs, as well
as for the magnetic structure of the contact, has not yet been included into
the circuit theory. We have investigated this problem in some details \cite
{huertas2}. Here, we only give the results for the relevant case of a tunnel
connector:

\begin{eqnarray}
\check{I}_{21} &=&\frac{G_{\text{T}}}{2}\left[ \check{G}_{2},\check{G}_{1}%
\right] +\frac{G_{\text{MR}}}{4}\left[ \{\vec{M}\hat{\vec{\sigma}}\hat{\tau}%
_{3},\check{G}_{2}\},\check{G}_{1}\right]  \nonumber \\
&&+i\frac{G_{\phi }}{2}\left[ \vec{M}\hat{\vec{\sigma}}\hat{\tau}_{3},\check{%
G}_{1}\right] \,.  \label{I_current}
\end{eqnarray}
Here $\check{G}_{1(2)}$ are the Green's functions on the two sides of the
junction. They are matrices in Keldysh-Nambu-spin space, obeying the
normalization condition $\check{G}^{2}=\check{1}$\cite{RefRammer}. The first
term presents the usual boundary condition for tunnel junctions \cite
{RefYuli}, $G_{T}$ being the junction conductance. The second term accounts
for the different conductances for different spin directions. This term
leads to a spin polarized current through the junction. We assume a small
value of this effect, $G_{\text{MR}}\sim G_{\text{T}}^{\uparrow }-G_{\text{T}%
}^{\downarrow }\ll G_{\text{T}}$. The unity vector $\vec{M}$ is in the
direction of the magnetization, and $\hat{\vec{\sigma}}$, $\hat{\tau}$ are
Pauli matrices in spin and Nambu space, respectively.

The third term is of the most interest for us. It will not vanish even if
there is no conductance through the junction. In this special case, the
physical meaning of the third term can be understood as follows: electrons
with different spin directions pick up different phases when reflecting from
the magnetic insulator. The coefficient $G_{\phi }$ is related to the mixing
conductance introduced in \cite{RefArne} via $G_{\phi }=\text{Im}G^{\uparrow
\downarrow }$. To give a concrete example, we have calculated $G_{\phi }$ in
the framework of an effective mass model for electrons with Fermi momentum $%
k_{\text{f}}$ and with spin-dependent penetration depths $\kappa _{\uparrow
,\downarrow }^{-1}$ \cite{RefTokuyasu}. Assuming $\delta \kappa \equiv
\kappa _{\uparrow }-\kappa _{\downarrow }\ll \kappa $, we find $G_{\phi
}=16AG_{\text{Q}}\kappa \delta \kappa \text{ arcsin}\left( ik_{\text{f}%
}/\kappa \right) /(k_{\text{f}}^{2}+\kappa ^{2})$, $A$ being the surface
area of the interface and $G_{Q}\equiv e^{2}/2\pi \hbar $. First principles
calculations of these interface spin conductances have been performed
recently \cite{RefXia}.

We proceed by finding the Green's functions for equilibrium conditions. In
particular it is sufficient to find the solution in the retarded block only.
The retarded Green's functions associated with the ferromagnetic and with
the superconducting reservoirs are respectively $\hat{R}_{\text{F}}=\hat{\tau%
}_{3}$ and $\hat{R}_{\text{S}}=\hat{\tau}_{1}$, assuming that the range of
energies considered is smaller than gap of the superconducting reservoir ($%
\varepsilon \ll \Delta _{\text{bulk}}$). The retarded function $\hat{R}$ in
the normal metal is obtained from the conservation of matrix currents in the
node. The current from the superconductor is given by the first term in (\ref
{I_current}), and the current from the ferromagnetic insulator is given by
the third term. A further current (called ``leakage current'' in Ref.~%
\onlinecite{RefYuli}) being proportional to energy $\varepsilon $ and
inversely proportional to the average level spacing $\delta $ in the normal
node, is also included. It describes decoherence between electrons and
holes. The matrix current conservation then reads 
\begin{equation}
\left[ -iG_{\text{Q}}\frac{\varepsilon }{\delta }\hat{\tau}_{3}-i\frac{%
G_{\phi }}{2}\vec{M}\hat{\vec{\sigma}}\hat{\tau}_{3}+\frac{G_{\text{T}}^{(%
\text{S})}}{2}\hat{\tau}_{1}\,,\,\hat{R}\right] =0\,,  \label{Commut1}
\end{equation}
where $G_{\text{T}}^{(\text{S})}$ is the conductance of the tunnel junction
to the superconductor. This equation is easy to solve since it again
separates into two blocks for spin parallel $(\uparrow )$ and antiparallel $%
(\downarrow )$ to the magnetization. We introduce parameters $\tilde{h}%
\equiv G_{\phi }\delta /2G_{Q}$ and $\tilde{\Delta}=G_{T}^{(S)}\delta /2G_{Q}
$. In these notations, the normalized density of states in the normal node
is different for two spin directions and reads 
\begin{equation}
\upsilon ^{\uparrow (\downarrow )}(\varepsilon )=\frac{\left| \varepsilon
\pm \tilde{h}\right| }{\sqrt{(\varepsilon \pm \tilde{h})^{2}-\tilde{\Delta}%
^{2}}}  \label{DOS}
\end{equation}
This expression is the same as the one for a BCS superconductor in the
presence of the spin-splitting magnetic field\cite{RefTedrow}. However, here
the density of states is formed in the normal metal, where neither
superconductivity nor magnetization are present. The quantities $\tilde{%
\Delta}$, $\tilde{h}$ are {\em induced} by the corresponding reservoirs.
This is why superconductivity and ferromagnetism do not have to compete and
the relevant parameters can be experimentally controlled by adjusting the
conductivities of the barriers \cite{comment}.

Having obtained the simple solution (\ref{DOS}), we discuss now the limits
of its validity. The first limitation is the presence of sufficiently strong
scattering in the normal part and/or at its boundaries to provide the
isotropy of the Green function. Two other limitations are provided by the
homogeneity of the Green's function in the node. The minimum size $L$ of the
normal part should exceed neither the superconducting coherence length nor
the spin-flip length. If the size of the system is larger than the spin-flip
length, the circuit theory description fails and spatially dependent Green's
functions have to be considered. In addition, the conductance of the normal
part itself should exceed both $G_{\phi }$ and $G_{\text{T}}^{\text{(S)}}$.

Now we consider transport between the two S/N/F structures through a
non-magnetic tunnel junction with conductance $G_{\text{T}}^{\text{(J)}}$
connecting the two normal metals (see Fig.1). Both structures are assumed to
be in local equilibrium. This assumption is justified if $G_{\text{T}}^{%
\text{(J)}}\ll G_{\phi },$ $G_{\text{T}}^{\text{(S)}}$. A voltage $V$ \ is
applied between them. We also assume that the temperature $T$ is much
smaller than $\tilde{\Delta},$ $\tilde{h}$. This is required for the
absolute spin-valve effect. The magnetization directions $\vec{M}_{1(2)}$ of
each magnetic insulator may be arbitrary. The matrix current between the two
nodes reads 
\begin{equation}
\check{I}=\frac{G_{\text{T}}^{\text{(J)}}}{2}\left[ \check{G}_{1},\check{G}%
_{2}\right] \,,  \label{I_commut}
\end{equation}
where $\check{G}_{1(2)}$ are the quasiclassical Green's functions for the
left $(1)$ and for the right node $(2)$ respectively. We can choose the
spin-quantization axis to be parallel to $\vec{M}_{1}$. As a result, the
Green's function $\check{G}_{1}$ separates into two blocks in spin space 
\begin{equation}
\check{G}_{1}=\left[ 
\begin{array}{cc}
\check{G}_{1}^{\upuparrows } & 0 \\ 
0 & \check{G}_{1}^{\downdownarrows }
\end{array}
\right] \,.  \label{GL}
\end{equation}
The Green's function $\check{G}_{2}$ can be presented as 
\begin{equation}
\check{G}_{2}=U\;\left[ 
\begin{array}{cc}
\check{G}_{2}^{\upuparrows } & 0 \\ 
0 & \check{G}_{2}^{\downdownarrows }
\end{array}
\right] \;U^{-1}\,.  \label{GR}
\end{equation}
$U$ $\ $being the spin rotation matrix that transforms $\vec{M}_{2}$ into $%
\vec{M}_{1}$. The electric current is given by the Keldysh component of Eq.~(%
\ref{I_commut}) 
\begin{equation}
I_{e}=-\frac{G_{\text{T}}^{(J)}}{8\text{ }e}\int_{-\infty }^{\infty
}d\varepsilon \text{Tr}\left\{ \hat{\tau}_{3}\left[ \check{G}_{1},\check{G}%
_{2}\right] ^{\text{K}}\right\} \,.  \label{I_e}
\end{equation}
It may be written as 
\begin{equation}
I_{\text{e}}=\frac{1}{4}I_{\text{p}}+\frac{1}{4}I_{\theta }(\vec{M}_{1}\cdot 
\vec{M}_{2}),  \label{Ie}
\end{equation}
where $I_{\text{p},\theta }=I^{\uparrow \uparrow }+I^{\downarrow \downarrow
}\pm I^{\uparrow \downarrow }\pm I^{\downarrow \uparrow }$. Each $%
I^{ss^{\prime }}$ ( $s$ and $s^{\prime }=\left\{ \uparrow ,\downarrow
\right\} $) is an integral of the form 
\begin{equation}
I^{ss^{\prime }}=\frac{G_{\text{T}}^{(J)}}{e}\int_{0}^{eV}d\varepsilon \text{
\ }\upsilon _{1}^{s}(\varepsilon -eV)\text{ }\upsilon _{2}^{s^{\prime
}}(\varepsilon )\,  \label{I_current_componets}
\end{equation}
\cite{Refcomment1}. As a function of the applied bias voltage, the left
density of states $\upsilon _{1}^{s}(\varepsilon -eV)$ is shifted in energy.
Now we assume  $\tilde{\Delta}_{1(2)}\geqslant \tilde{h}_{1(2)}$. Each
component $I^{ss^{\prime }}$ will be zero until the voltage reaches a
certain threshold $eV_{\text{th}}^{ss^{\prime }}$, at which both left and
right densities of states start to overlap. Because both densities of states
are spin-split, there are four different threshold voltage $eV_{\text{th}%
}^{ss^{\prime }}$, depending on which spin components of both densities of
states are ``matched'' together 
\begin{equation}
eV_{\text{th}}^{ss^{\prime }}=\tilde{\Delta}_{1}+\tilde{\Delta}_{2}\mp
\left( \tilde{h}_{1}\pm \tilde{h}_{2}\right) .  \label{eVss}
\end{equation}
So the voltage interval $|eV-\tilde{\Delta}_{1}-\tilde{\Delta}_{2}|\leqslant 
\tilde{h}_{1}+\tilde{h}_{2}$ can be divided in four regions, separated by
the four different threshold voltages $eV_{th}^{ss^{\prime }}$.

To illustrate the effect, we consider the symmetric case{\em \ } $\tilde{%
\Delta}_{1}=\tilde{\Delta}_{2}\equiv \tilde{\Delta}$, $\tilde{h}_{1}=\tilde{h%
}_{2}\equiv \tilde{h}$. In this case, there are only three threshold
voltages $eV_{\text{th}}^{\downarrow \text{ }\uparrow }=2\left( \tilde{\Delta%
}-\tilde{h}\right) ,eV_{\text{th}}^{\uparrow \text{ }\uparrow \text{ (}%
\downarrow \text{ }\downarrow \text{)}}=2\tilde{\Delta}$ and $eV_{\text{th}%
}^{\uparrow \text{ }\downarrow }=2\left( \tilde{\Delta}+\tilde{h}\right) $.
At each threshold, the correspondent spin component $I^{ss^{\prime }}$ jumps
from zero to the value 
\begin{equation}
I^{ss^{\prime }}\approx \frac{\pi }{4}\frac{G_{\text{T}}^{(J)}}{e}eV_{\text{%
th}}^{ss^{\prime }}.  \label{Iss}
\end{equation}
These jumps are characteristic of tunneling between superconductors (S-S
tunneling) \cite{RefTinkham}. Through the voltage interval $|eV-2\tilde{%
\Delta}|\leqslant 2\tilde{h}$, the total current $I_{\text{e}}$ presents
steps reflecting these jumps (Fig. 2). These steps depend on the relative
angle $\theta $ between the magnetization of the magnetic insulators (see
Fig. 2). Of specific interest is the first jump of the current in
antiparallel configuration ($\theta =\pi $), occurring at the threshold $eV_{%
\text{th}}^{\downarrow \text{ }\uparrow }=2\left( \tilde{\Delta}-\tilde{h}%
\right) $. In this case only spin-down quasiparticles in the left node
overlap with spin-up quasiparticles in the right node, which constitutes the 
{\em absolute spin valve-effect}. As expected, the total current being
finite at $\theta =\pi $, goes to zero if the magnetization of one of the
ferromagnetic insulator is reversed (see Fig.2). The {\em absolute spin
valve-effect} already vanishes at the second zone. Nevertheless again the
difference between $\theta =\pi $ and $\theta =0$ currents resembles the
effect. Generally these results depend on the relative values of $\tilde{h}%
_{1(2)}$ and $\tilde{\Delta}_{1(2)}$. In general the region of voltages
where the effect occurs $eV_{\text{th}}^{\downarrow \text{ }\downarrow }-eV_{%
\text{th}}^{\downarrow \text{ }\uparrow }$ is equal to $2$ {\em min(}$\tilde{%
h}_{1}$, $\tilde{h}_{2}${\em )}.

In conclusion, we have investigated theoretically spin transport in
multiterminal S/N/F proximity structures using quasiclassical Green's
function methods, inspired by circuit theories of mesoscopic transport \cite
{RefYuli,RefArne}. Spin-splitting of the induced density of states, caused
by the presence of magnetic insulators, is probed by means of tunneling
spectroscopy of the superconducting proximity effect. The tunneling current
has jumps for certain intervals of voltages, in which an {\em absolute spin
valve effect} can be achieved. These features of the current depend on the
relative angular configuration of the different magnetic insulators and on
the relative values of the induced superconducting minigap and the induced
spin-splitting in each node. Moreover our proposal allows for the
possibility of inducing two independent ``fields'' (i.e. antiparallel
``fields'') in the device. This would be very difficult to achieve with an
applied magnetic field in a system of superconducting electrodes. Finally,
we emphasize, that the physical separation of the sources of both
superconducting and ferromagnetic correlations provides a feasible way to
manipulate specifically the spin-filtering properties of our proposed
multiterminal S/N/F proximity structure.

We thank discussions with A. Brataas, Ya. M. Blanter, Gerrit E. W. Bauer and
M.R. Wegewijs. We specially thank N.M. Chtchelkatchev for stimulating
discussions and for his help with the numerical calculations. This work was
financially supported by the Stichting voor Fundamenteel Onderzoek der
Materie (FOM).

\bigskip

\bigskip

\bigskip

\bigskip

\bigskip

\bigskip

\bigskip

\bigskip

\bigskip

\bigskip

\bigskip

\bigskip

\bigskip

\bigskip

\bigskip

\bigskip

\bigskip

\bigskip

\bigskip

FIG. 1. Schematic circuit of two coupled tri-layer S/N/F structures. In each
trilayer structure, a normal metal node (N) is coupled to superconducting
(S) and ferromagnetic (F) reservoirs through tunnel junctions of
conductances $G_{T}^{(S)}$ and $G_{\phi }$ respectively. The ferromagnetic
reservoir is assumed to be a magnetic insulator. Both normal metal nodes are
coupled together through a third tunnel junction $G_{\text{T}}^{(J)}$. The
relative magnetic configuration of the ferromagnetic insulators may be
non-collinear. A voltage $V$ is applied between both N nodes.

\bigskip

\bigskip

FIG. 2. Steps of the normalized N-N tunneling current $e$ $I_{\text{e}}/(G_{%
\text{T}}^{(J)}\Delta )$ with the applied voltage $V$ for the symmetric case%
{\em \ } $\tilde{\Delta}_{1}=\tilde{\Delta}_{2}\equiv \tilde{\Delta}$, $%
\tilde{h}_{1}=\tilde{h}_{2}\equiv \tilde{h}$. In this$\ $case $\tilde{h}/%
\tilde{\Delta}=$ $0.5$. The tunneling current presents jumps in the range of
voltages $|eV-2\tilde{\Delta}|\leqslant 2\tilde{h}$. For $\theta =0$, the
current jumps at the voltage $eV=2\tilde{\Delta}$. For $\theta =\pi $ the
current presents two jumps at voltages $eV=2(\tilde{\Delta}-\tilde{h})$ and $%
eV=2(\tilde{\Delta}+\tilde{h})$ respectively. This jumps reflect how the
different spin components of the induced density of states in each normal
node, contribute to the total tunneling current at different voltages.
Between the voltages $eV=2(\tilde{\Delta}-\tilde{h})$ and $eV=2\tilde{\Delta}
$ the {\em absolute spin-valve effect} is achieved. The change of the
current between $\theta =0$ and $\theta =\pi $ situations is shown for
various values of the angle $\theta $.


\begin{references}
\bibitem{comment}  In particular we found that the parameters that control
the injection of the ``ferromagnetic'' and ``superconducting'' correlations
are inversally proportional to the average time, that it takes for the
quasiparticles to probe each of the junction that connect the node with the
reservoirs ($\tau _{S}$ $=\hbar /\tilde{\Delta}$, and $\tau _{F}=\hbar /%
\tilde{h}$ respectively).
\bibitem{Refcomment1}  There is also a term $I_{\text{J}}$ in $I_{\text{e}}$%
, which describes supercurrent between the two nodes: $I_{\text{J}}(t)=I_{%
\text{J}}(eV)\sin (2eVt+\Phi )$. For a d.c. bias voltage, this Josephson
current alternates and does not contribute to d.c. electric current.
\end{references}
\end{document}